\documentclass[twocolumn]{aastex63}
\usepackage{subscript}
\usepackage{color}

\accepted{for publication in ApJ}
\shorttitle{NUMBER DENSITY PROFILE OF GLOBULAR CLUSTERS}
\shortauthors{Lee, Chung \& Yoon}
\begin{document}  
\title{NONLINEAR COLOR$-$METALLICITY RELATIONS OF GLOBULAR CLUSTERS. IX. DIFFERENT RADIAL NUMBER DENSITY PROFILES BETWEEN BLUE AND RED CLUSTERS}
\correspondingauthor{Suk-Jin Yoon}
\email{sjyoon0691@yonsei.ac.kr}
\author[0000-0002-7957-3877]{Sang-Yoon Lee}
\affiliation{Center for Galaxy Evolution Research, Yonsei University, Seoul 03722, Republic of Korea}
\author[0000-0001-6812-4542]{Chul Chung}
\affiliation{Center for Galaxy Evolution Research, Yonsei University, Seoul 03722, Republic of Korea}
\affiliation{Department of Astronomy, Yonsei University, Seoul 03722, Republic of Korea}
\author[0000-0002-1842-4325]{Suk-Jin Yoon}
\affiliation{Center for Galaxy Evolution Research, Yonsei University, Seoul 03722, Republic of Korea}
\affiliation{Department of Astronomy, Yonsei University, Seoul 03722, Republic of Korea}

\begin{abstract}
The optical colors of globular clusters (GCs) in most large early-type galaxies are bimodal. 
Blue and red GCs show a sharp difference in the radial profile of their surface number density in the sense that red GCs are more centrally concentrated than blue GCs.
An instant interpretation is that there exist two distinct GC subsystems having different radial distributions.
This view, however, was challenged by a scenario in which, due to the nonlinear nature of the GC metallicity-to-color transformation for old ($\gtrsim$\,10 Gyr) GCs, a broad unimodal metallicity spread can exhibit a bimodal color distribution.
Here we show, by simulating the radial trends in the GC color distributions of the four nearby giant elliptical galaxies (M87, M49, M60, and NGC 1399), that the difference in the radial profile between blue and red GCs stems naturally from the metallicity-to-color nonlinearity plus the well-known radial metallicity gradient of GC systems.
The model suggests no or little radial variation in GC age even out to $\sim$20\,${R}_{\rm eff}$.
Our results provide a simpler solution to the distinct radial profiles of blue and red GCs that does not necessarily invoke the presence of two GC subsystems and further fortify the nonlinearity scenario for the GC color bimodality phenomenon.
\end{abstract}
\keywords{Galaxy evolution (594); Giant elliptical galaxies (651); Elliptical galaxies (456); cD galaxies (209); Globular star clusters (656)}

\section{Introduction}
\label{introduction}

\begin{figure}[htbp]
\centering
\includegraphics[keepaspectratio,height=0.86\textheight]{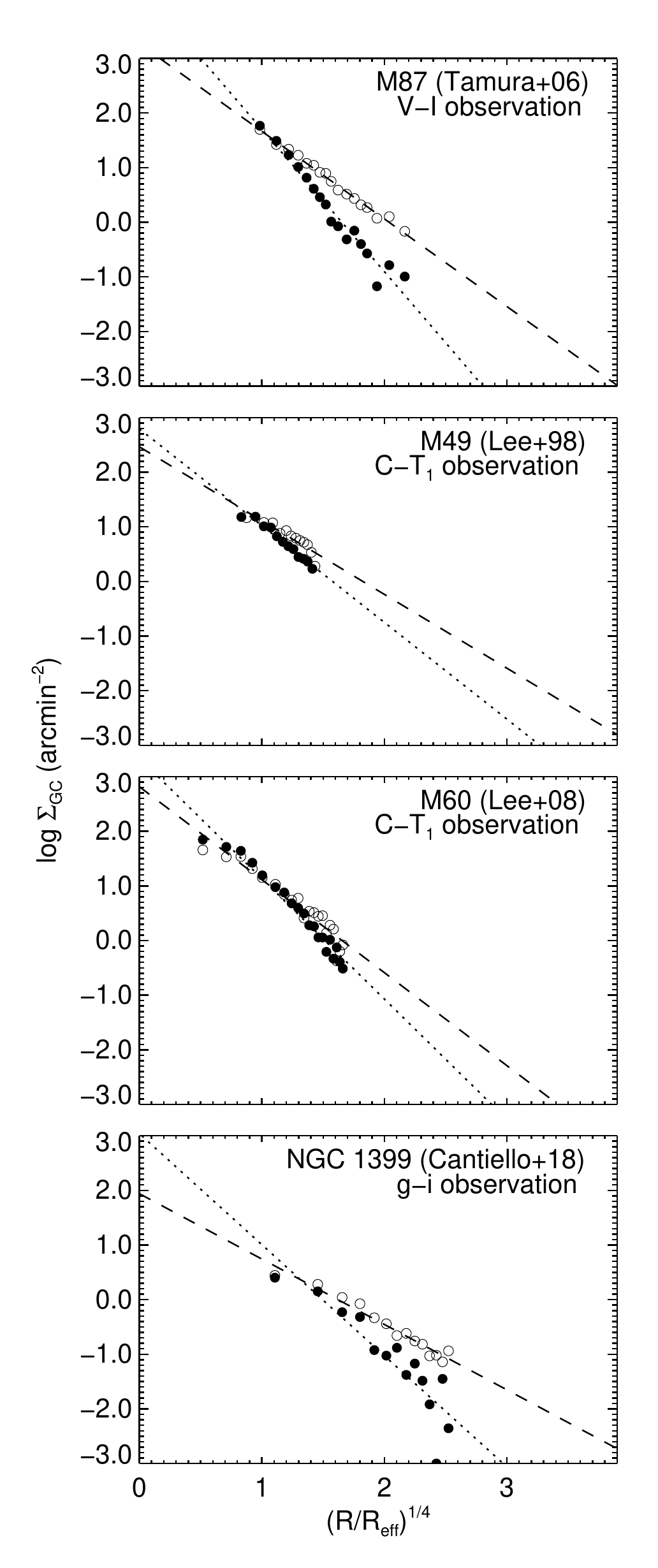}
\caption{\label{fig:Figure1}
Radial surface number density profiles of blue (open circles) and red (filled circles) GCs in the four giant elliptical galaxies. We use data from \citet{2006MNRAS.373..601T} for M87, \citet{1998AJ....115..947L} for M49, \citet{2008ApJ...682..135L} for M60, and \citet{cantiello18} for NGC 1399. The dashed and dotted lines are fitted to the blue and red GCs, respectively, following the de Vaucouleurs $R^{1/4}$ law.}
\end{figure}

Globular clusters (GCs) play a vital role in understanding the formation process of their host galaxies because GC formation accompanies star-forming episodes of their hosts, and GCs are easier to observe than field stars of their hosts.
Among others, a well-established observational phenomenon for GCs, the bimodal optical color distributions in early-type galaxies have shed light on galaxy and GC formation~\citep[e.g.,][to name a few; see also \citealt{2004Natur.427...31W,2006ARA&A..44..193B,2020rfma.book..245B} for reviews and references therein]{1993AJ....105.1762O,1993MNRAS.264..611Z,1999AJ....118.1526G,2001AJ....122.1251K,2001AJ....121.2974L,2005MNRAS.357...56F,2012MNRAS.421..635F,2006ApJ...636...90H,2017ApJ...835..101H,2006ApJ...639...95P,2008ApJ...674..857L,2010ApJ...710...51B,2011MNRAS.416..155F,2011MNRAS.413.2943F,2011MNRAS.415.3393F,2011ApJ...728..116L,2012MNRAS.420...37B,2012A&A...539A..54C,2012MNRAS.422.3591C,2013ApJ...763...40K,2013MNRAS.436.1172U,2014A&A...564L...3C,cantiello18,2014MNRAS.437..273K,2016MNRAS.458..105K,2015A&A...574A..21R,2016ApJ...822...95C,2017MNRAS.470.3227C,2018MNRAS.474.4302E,2020MNRAS.493.2253E,2018MNRAS.477.3869H,2019MNRAS.488..770E,2019ApJ...872..202K,2020MNRAS.492.4313D}.
The bimodal distribution of colors suggests the existence of two distinct GC subpopulations.
In order to explain two GC subpopulations within a galaxy, several galaxy formation scenarios have been put forward, including the merger~\citep{1972ApJ...178..623T,1992ApJ...384...50A,1993MNRAS.264..611Z,1995AJ....109..960W,1997AJ....114.2381M}, accretion~\citep{1987ApJ...313..112M,1998ApJ...501..554C,2002ApJ...567..853C,1999A&AS..138...55H}, and multiphase formation~\citep{1994ApJ...429..177H,1997AJ....113..887F,1999AJ....117..855H}. 
The more recent two-phase formation scenario for galaxies ~\citep{2011MNRAS.413.2943F,2013ApJ...773L..27P,2016ApJ...822...70L,2018Natur.555..483B} is in line with ``accretion."
These explanations were to fulfill the prerequisite that the color bimodality is a manifestation of the two GC subpopulations with distinct geneses.

In relation to the GC color bimodality, a well-known phenomenon observed in most GC systems is the systematic variation in the GC color histogram morphology along the galactocentric radius~\citep{1996AJ....111.1529G,1999ApJ...513..733K,2000AJ....120..260L,2003AJ....125.1908D,2005A&A...433...43D,2004MNRAS.355..608F,2011MNRAS.413.2943F,2006A&A...451..789B,2006MNRAS.373..601T,2011ApJS..197...33S,2012MNRAS.420...37B,2013ApJ...763...40K,2015MNRAS.449..612E,2018MNRAS.474.4302E,2016ApJ...817...58H,2019MNRAS.488..770E,2019ApJ...872..202K,2020MNRAS.492.4313D}.
For most early-type galaxies, the fraction of red (blue) GCs is highest (lowest) at the galactic center and decreases (increases) with radius.
Accordingly, the surface number density of red GCs drops faster than that of blue GCs as the radius increases~\citep{2004MNRAS.355..608F,2005MNRAS.357...56F,2006A&A...451..789B,2008MNRAS.386.1145B,2006MNRAS.373..601T,2009ApJ...703..939H,2011MNRAS.416..155F,2011ApJS..197...33S,2012MNRAS.420...37B,2013MNRAS.428..389P,2013MNRAS.436.1172U,2016ApJ...822...95C,2016ApJ...817...58H,2017ApJ...835..101H,2018MNRAS.474.4302E,2019MNRAS.488..770E,2019ApJ...872..202K,2020MNRAS.492.4313D}.
In Figure~\ref{fig:Figure1}, we show the surface number density profiles of blue and red GCs in the four largest elliptical galaxies (M87, M49, M60, and NGC 1399) in the Virgo and Fornax galaxy clusters.
The blue and red GCs show a sharp difference in the radial profile in the sense that the red GCs are more centrally concentrated than blue GCs.
This has been generally interpreted as the presence of two subpopulations having distinct spatial occupations.

\begin{table*}
\footnotesize
\begin{center}
\caption{Observational Data\label{tab:Table1}}
\begin{tabular}{lccccccccccl}
\tableline
\tableline
Galaxy      && Telescope &  & Filter & & $N_{\rm GC}$$^a$      &  & Radial Extension (kpc)$^b$  & Radial Extension ($R_{\rm eff}$) & & Reference\\
\tableline
M87         && HST       &  & $gz$   & &  1745                 &  &  0.1$-$13.1 &  0.02$-$1.71    & & \citet{2009ApJS..180...54J} \\
(NGC 4486)  && HST       &  & $VI$   & &  2250                 &  &  0.1$-$12.2 &  0.01$-$1.58    & & \citet{2009ApJ...703...42P} \\
            && KPNO/CTIO &  & $CT_1$ & &  2933                 &  &  3.1$-$61.0 &  0.40$-$7.93    & & \citet{2007MNRAS.382.1947F} \\
\tableline                                                                                      
M49         && HST       &  & $gz$   & &  765                  &  &  0.3$-$12.3 &  0.03$-$1.46    & & \citet{2009ApJS..180...54J} \\
(NGC 4472)  && HST       &  & $VI$   & &  609                  &  &  0.0$-$18.6 &  0.01$-$2.21    & & \citet{2000AJ....120..260L} \\
            && KPNO      &  & $CT_1$ & &  1999                 &  &  1.9$-$48.4 &  0.22$-$5.74    & & \citet{2006ApJ...647..276K} \\
\tableline                                                                                      
M60         && HST       &  & $gz$   & &  807                  &  &  0.1$-$13.4 &  0.02$-$2.43    & & \citet{2009ApJS..180...54J} \\
(NGC 4649)  && Gemini    &  & $g^{\prime}i^{\prime}$ & &  1546 &  &  0.8$-$43.4 &  0.15$-$7.86    & & \citet{2011MNRAS.416..155F} \\
            && KPNO      &  & $CT_1$ & &  1539                 &  &  4.8$-$50.7 &  0.88$-$9.18    & & \citet{2008ApJ...682..135L} \\
\tableline                                                                                      
NGC 1399    && HST       &  & $gz$   & &  1075                 &  &  0.4$-$15.6 &  0.09$-$3.86    & & \citet{2015ApJS..221...13J} \\
            && CTIO      &  & $BI$   & &  2037                 &  &  4.2$-$196.3 &  1.05$-$48.42  & & \citet{2013ApJ...763...40K} \\
            && KPNO/CTIO &  & $CT_1$ & &  1920                 &  &  3.1$-$65.6 &  0.76$-$16.19   & & \citet{2007MNRAS.382.1947F} \\
\tableline
\end{tabular}
\end{center}
\tablecomments{The radial extension values correspond to the innermost and the outermost GCs in each catalog.}
\tablenotetext{a}{The number of GCs in the catalog.}
\tablenotetext{b}{We take the distance to each galaxy from \citet{2009ApJ...694..556B}.}
\end{table*}

As opposed to the popular belief in the existence of GC subpopulations with different metallicities, \citet[][hereafter Paper I; see also \citealt{2006BASI...34...83R}]{2006Sci...311.1129Y} offered an alternative explanation, in which the primary driver of color bimodality is the nonlinear nature of the color$-$metallicity relations (CMRs).
They showed that the CMRs of old ($>$10 Gyr) GC systems are inflected due mainly to helium-burning horizontal branch (HB) stars, and that the color bimodality can be naturally achieved from broad, unimodal metallicity distributions by the nonlinear CMRs.
In a series of subsequent papers, the nonlinearity scenario was tested both observationally and theoretically.
\citet[][hereafter Paper II]{2011ApJ...743..149Y} and \citet[][hereafter Paper IV]{2013ApJ...768..137Y} revealed using M87 (Paper II) and M84 (Paper IV) GC systems that the degree of the CMR nonlinearity depends on the choice of colors ($g-z$, $u-z$, and $u-g$) and governs the shape of color histograms.
\citet[][hereafter Paper III]{2011ApJ...743..150Y} demonstrated that the shape of the metallicity distribution functions (MDFs) of GC systems are characterized by a broad, skewed Gaussian function and are similar to MDFs of both halo field stars and galactic chemical evolution models, alleviating the long-standing discrepancy between GC and stellar MDFs.
\citet[][hereafter Paper VIII]{2019ApJS..240....2L} reproduced the GC color distributions of 78 early-type galaxies in the Virgo and Fornax galaxy clusters.
They showed that $\sim$\,$70\,\%$ of the GC systems fit into the nonlinearity theory and the remaining $\sim$\,$30\,\%$ are also consistent with the theory assuming a young GC population diluting the underlying color bimodality.
The nonlinearity scenario was further explored by using spectroscopic absorption-line index distributions as a close analogy to the photometric color distributions~\citep[][Papers V, VI, and VII]{2013ApJ...768..138K,2016ApJ...818..201C,2017ApJ...843...43K}.
They showed that the inflected index$-$metallicity relations of GCs account for the observed {\it bimodal} absorption-line index distributions of M31 GCs (Paper V) and NGC 5128 GCs (Paper VII).
In the same vein, Paper VI revisited the well-known metallicity proxy, Ca II triplet (CaT) index, and showed that the inflected CaT$-$metallicity relation is responsible for the observed CaT bimodality of GCs in elliptical galaxies. 
Most recently, \citet[][hereafter Paper X]{Kim20} obtained the spectroscopic metallicities of $\sim$\,130 GCs of M87 with Subaru/FOCAS and confirmed that the CMRs are nonlinear and the MDF is close to a unimodal distribution.

According to the nonlinearity theory, {\it neither} the radial variation in the color distribution morphology (i.e., the relative portions of blue and red GCs) {\it nor} the radial difference in the density profiles of blue and red GCs is caused by two GC subpopulations occupying different spatial positions.
But instead the both radial behaviors are originated by a combined effect of the nonlinear CMRs plus the systematic change of the mean GC metallicity with galactocentric radius.
In this paper, we put our hypothesis to the test on the origin of the different number density profiles of blue and red GCs in giant elliptical galaxies.
This paper is organized as follows.
Section~\ref{sec:OBSdata} describes the observational database used in this study.
Section~\ref{sec:simulations} gives descriptions of the stellar population model and the GC color distribution model.
Section~\ref{sec:results} compares between the observations and our simulations in terms of the radial variations in the color distribution morphology and the difference in the surface number density profiles of blue and red GCs.
In Section~\ref{sec:discussion}, we discuss the implications of our results.

\vspace{1mm}
\section{Observational Data}
\label{sec:OBSdata}

\begin{figure*}[htbp]
\centering
\includegraphics[keepaspectratio,width=\textwidth,height=0.75\textheight]{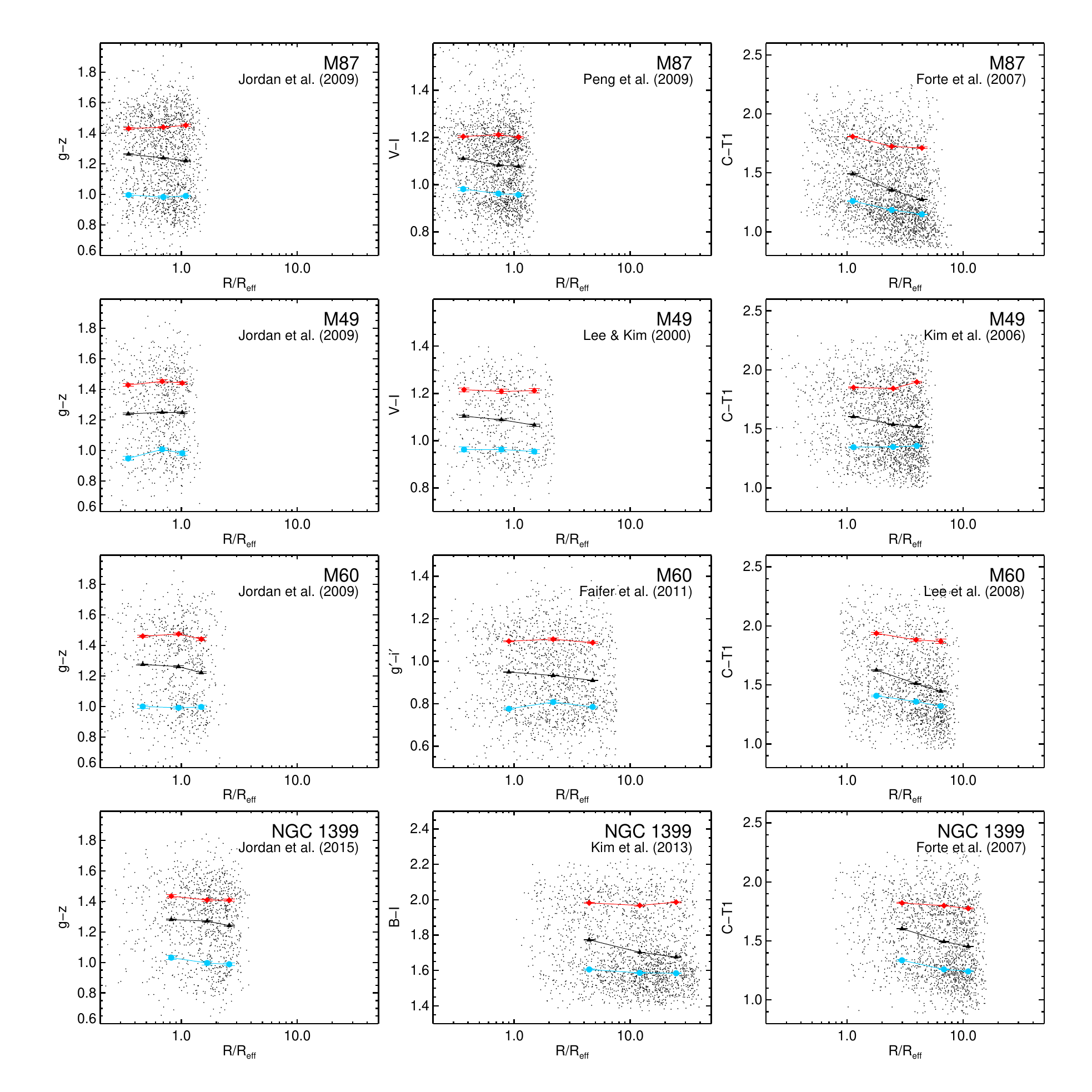}
\caption{\label{fig:Figure2}
Colors of individual GCs against the radii for three different observations for each galaxy.
The blue and red lines represent the mean color of blue and red GCs that are determined by the KMM code in each radial bin.
The black lines indicate the mean colors of the entire GCs.
The error bars were obtained by carrying out 1000 bootstrapping iterations.}
\end{figure*}

\begin{figure*}[htbp]
\centering
\includegraphics[keepaspectratio,width=\textwidth,height=0.75\textheight]{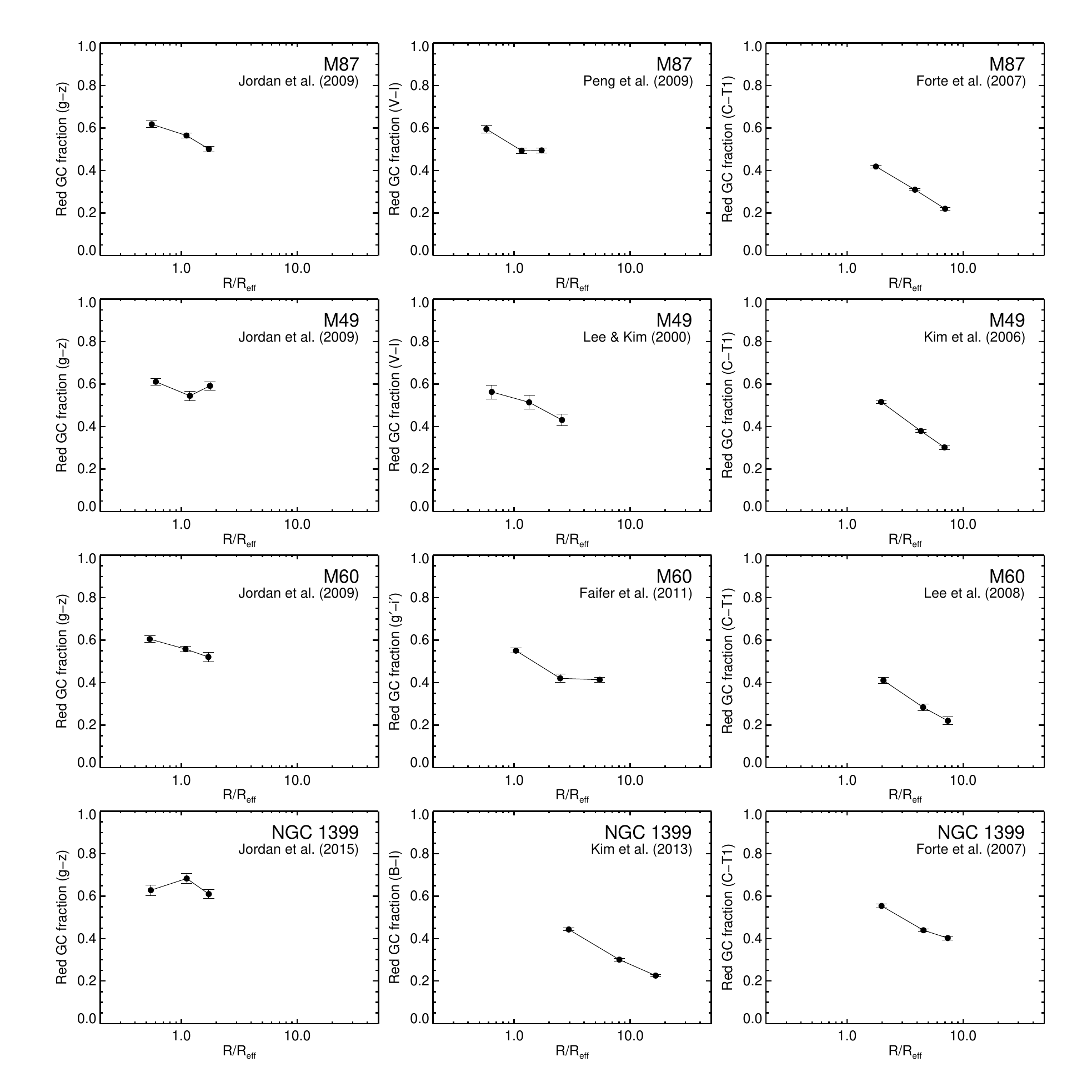}
\caption{\label{fig:Figure3}
Same as Figure~\ref{fig:Figure2}, but for the number fractions of the red GCs for each galaxy.}
\end{figure*}

\subsection{Information on Individual GC Systems}
\label{InformationOnIndividualGCSystems}

We analyze the four largest giant elliptical galaxies\footnote{We refer to the HyperLeda database (http://leda.univ-lyon1.fr/) for the morphological classification and \citet{2011ApJ...728..116L} for the stellar mass of galaxies.} (M87, M49, M60, and NGC 1399) in the Virgo and Fornax galaxy clusters. 
They harbor numerous GCs, for which we can examine the radial properties.
The GC systems of the four galaxies have been observed frequently, and each galaxy has several photometric catalogs.
For each galaxy, we choose three GC catalogs with more than 500 GCs, for which color information along the radial direction is publicly available.
Table~\ref{tab:Table1} summarizes the photometric GC datasets used in this study.

In Figure~\ref{fig:Figure2}, we show the colors of individual GCs against their galactocentric radii.
The criterion for dividing radial bins is to assign an equal number of GCs to each bin.
The mean colors of blue and red GCs and the red fractions are determined by the KMM code by~\citet{1994AJ....108.2348A}.
The mean colors of blue and red GCs tend to be bluer as the radial distance increases, except in a few cases (e.g., $g-z$ for M49).
In Figure~\ref{fig:Figure3}, we show the red GC fractions of our sample galaxies as a function of the galactocentric radius.
As is well known, the number fraction of red GCs declines with radius~\citep[e.g.,][]{2006ApJ...636...90H,2017ApJ...835..101H,2013ApJ...763...40K}.
Thus, back in Figure~\ref{fig:Figure2}, the mean colors of the entire (blue + red) GCs (black lines) show steeper decline than the mean colors of blue GCs (blue lines) or red GCs (red lines).
Brief information on each galaxy's observational data sets is given below.

\begin{table}
\footnotesize
\begin{center}
\caption{The Empirical Color--Metallicity Relations\label{tab:Table2}}
\begin{tabular}{lcccccl}
\tableline
\tableline
 \multicolumn{7}{c}{[Fe/H]$=\alpha+\beta(X)+\gamma(X)^2$}\\
 \tableline
   $X$ & & $\alpha$ & $\beta$ & $\gamma$ & & Reference \\
\tableline
  \textit{$B-I$}    & & $-$5.94 &  2.79  &        0& & \cite{2000AJ....119..727B} \\
  \textit{$V-I$}    & & $-$5.39 &  4.22  &        0& & \cite{2000AJ....119..727B}  \\
  \textit{$g^{\prime}-i^{\prime}$}    & & $-$4.13 &  3.51  &        0& &\cite{2011MNRAS.416..155F}  \\
  \textit{$g-z$}    & & $-$4.02&  2.68   &        0& &\cite{2011ApJ...743..150Y}  \\
  \textit{$C-T_1$}  & & $-$6.04&  4.95   &   $-$0.98& &\cite{2002AJ....123.3108H}\\
\tableline                                                                  
\end{tabular}
\end{center}
\tablecomments{The ($g^{\prime}-i^{\prime}$)$-$[Fe/H] relation is obtained from the original ($g^{\prime}-i^{\prime}$)$-$[Z/H] relation~\citep{2011MNRAS.416..155F}, using the equation [Fe/H] = log(Z/Z$_\sun$)$-$\,log(X/X$_\sun$)$-$0.723\,[$\alpha$/Fe] = [Z/H]$-$0.217, with [$\alpha$/Fe] = 0.3~\citep{2002ApJS..143..499K}.}   
\end{table}

\subsubsection{M87 (NGC 4486)}
\label{m87}

$g-z$: the Advanced Camera for Surveys (ACS) Virgo Cluster Survey presents a catalog of GC candidates for 100 early-type galaxies~\citep{2009ApJS..180...54J}.
The imaging was done by ACS on board the Hubble Space Telescope (HST), and 12,763 GC candidates in $g-z$ were obtained in the whole galaxy sample. 
The field of view of ACS ($202\arcsec\times202\arcsec$) allows gradient measurements for the radius of $\lesssim$ 11.5 kpc at the Virgo distance of 16.7 Mpc~\citep{2007ApJ...655..144M,2011ApJ...728..116L}.
The number of GC candidates in M87 from this catalog is 1745.

$V-I$: \citet{2009ApJ...703...42P} constructed a photometric catalog of M87 GCs with F606W ($V_{606}$) and F814W ($I_{814}$) filters using ACS Wide Field Channel.
The radial coverage of this observation is similar to that of \citet{2009ApJS..180...54J}. 
The total number of GC candidates in this catalog is 2250.

$C-T_1$: \citet{2007MNRAS.382.1947F} presented a Washington $C$ and $T_1$ photometric catalog of the GC system in M87 and NGC 1399.
The photometric data were obtained by 4\,m telescopes at Kitt Peak National Observatory (KPNO) and Cerro Tololo Inter-American Observatory (CTIO). 
The radial coverage of this data is bigger than those of the two previous datasets.
They defined GC candidates, which satisfy the color and magnitude criteria ($0.9 < C-T_1 < 2.3$ and $21.0 < T_1 < 23.2$) among the photometric sources.
The number of GC candidates in M87 from this catalog is 2933.

\subsubsection{M49 (NGC 4472)}
\label{m49}

$g-z$: the data are taken from the ACS Virgo Cluster Survey catalog like M87 and M60.
The number of GC candidates in M49 is 765.

$V-I$: the observational data were obtained from HST/WFPC2 \citep{2000AJ....120..260L}.
This observation was performed for the central region ($r < 4\arcmin$) of the galaxy.
The color and magnitude criteria to select GC candidates are $0.75<V-I<1.45$ and $V$ $\leq$ 23.9.
A total of 609 GC candidates were in the catalog, and the sample is considered to be complete at $V\leq23.9$.

$C-T_1$: \citet{1996AJ....111.1529G} and their follow-up work, \citet{1998AJ....115..947L}, and \citet{2006ApJ...647..276K}, presented a Washington $C$ and $T_1$ photometric catalog of the GC system in M49 using the KPNO 4\,m telescope.
The color and magnitude criteria to select GC candidates are 1.0 $\leq$ $C-T_1$ $\leq$ 2.3 and 19.63 $\leq$ $T_1<23.0$~\citep{1996AJ....111.1529G}. 
The total number of GC candidates is 1999 at $r$ $\lesssim$ 9$\arcmin$.

\subsubsection{M60 (NGC 4649)}
\label{m60}

$g-z$: the data are taken from the ACS Virgo Cluster Survey catalog like M87 and M49.
The number of GC candidates is 807.

$g^{\prime}-i^{\prime}$: \citet{2011MNRAS.416..155F} presented $g^{\prime}$, $r^{\prime}$, and $i^{\prime}$ imaging with the Gemini North and South telescopes for the GC system in M60.
The photometric data on total 1546 GC candidates were obtained by the criteria of three-color selection (0.4 $\leq$ $g^{\prime}-i^{\prime}$ $\leq$ 1.45, 0.35 $\leq$ $g^{\prime}-r^{\prime}$ $\leq$ 0.95, and 0.0 $\leq$ $r^{\prime}-i^{\prime}$ $\leq$ 0.60) and 19.5 $\leq$ $i^{\prime}$ $\leq$ 24.3. 
The source list is complete by $\gtrsim$\,50\,\% at a magnitude of $i^{\prime}$ = 24.30.

$C-T_1$: \citet{2008ApJ...682..135L} presented a photometric catalog of the GCs in M60, based on wide-field Washington $CT_1$ images.
The imaging was done by the KPNO 4\,m telescope and completeness is higher than 90\,\% for $T_1 = 22.8$.
The observation covered $16\arcmin.4\times16\arcmin.4$, which is the most extensive field for M60.
We choose the GCs by the criteria of $1.0<C-T_1<2.4$, $19.0<T_1<23.0$, and $r>1\arcmin$, which were suggested by \citet{2008ApJ...682..135L}.
The final number of GC candidates in our sample is 1539.

\subsubsection{NGC 1399}
\label{ngc1399}

$g-z$: the data are taken from the ACS Fornax Cluster Survey~\citep{2015ApJS..221...13J}, and the criteria to select bona fide GCs are the same as the ACS Virgo Cluster Survey catalog.
The total number of GCs is 1075.

$B-I$: \citet{2013ApJ...763...40K} presented multiband ($U$, $B$, $V$, and $I$) photometry for NGC 1399 GCs with the CTIO 4\,m telescope.
The field of view ($36\arcmin\times36\arcmin$) of their observation covered almost the entire GC system in this galaxy.
The total number of GC is 2037 with $U$ $<$ 22.0.

$C-T_1$: the data are taken from \citet{2007MNRAS.382.1947F}, and the color and magnitude criteria are the same as the M87 case.
The total number of GCs is 1920.

\vspace{0.5cm}

\subsection{Radial Metallicity Gradients of GC Systems Derived from Empirical Color--Metallicity Relations}
\label{metallicitygradientsofgcsystemsderivedfromempiricalcolor$-$metallicityrelations}

In this study, we adopt the empirical CMRs to determine the mean [Fe/H] of the GC systems of our sample galaxies.
Table~\ref{tab:Table2} summarizes the adopted empirical CMRs and their references.

\begin{figure*}[htbp]
\centering
\includegraphics[keepaspectratio,height=0.38\textheight]{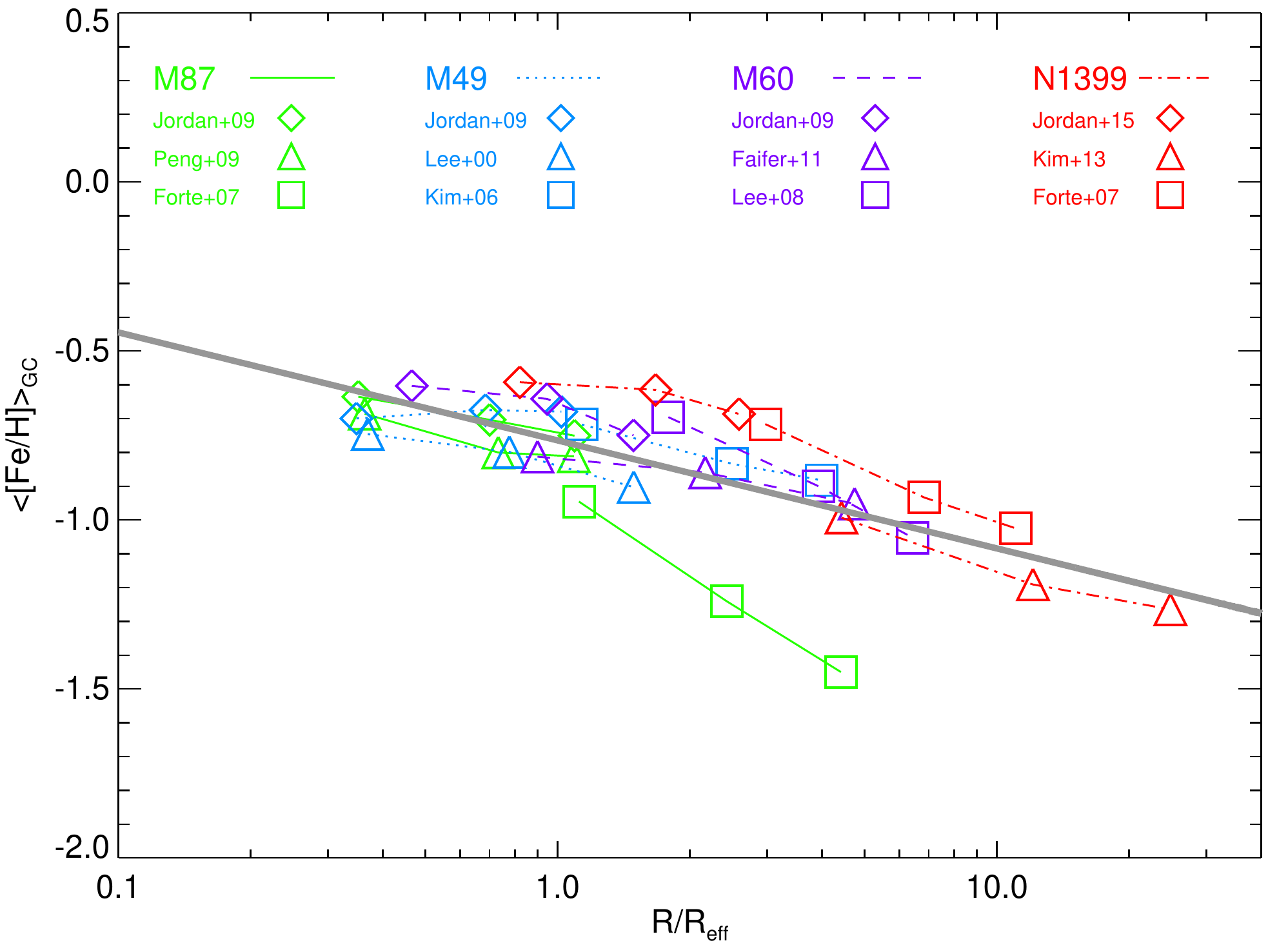}
\caption{\label{fig:Figure4}
Radial variation of ${\langle{\rm[Fe/H]}\rangle_{\rm GC}}$ of our sample GC systems. ${\langle{\rm[Fe/H]}\rangle_{\rm GC}}$ are derived from empirical CMRs. The radius for each galaxy is normalized by its effective radius. The gray solid line shows the least-squares fit to the combined data.}
\end{figure*}

In Figure~\ref{fig:Figure4}, we present the radial metallicity gradients of GC systems derived from the empirical CMRs.
The effective radius (${R}_{\rm eff}$) is used as a unit of radii to make a comparison among the four different galaxies.
The ${R}_{\rm eff}$ values of galaxies are obtained from the Third Reference Catalog of Bright Galaxies (RC3\footnote{\url{https://heasarc.gsfc.nasa.gov/W3Browse/all/rc3.html}};~\citealt{1991rc3..book.....D}).
The metallicities of the GC systems in the four galaxies decrease with radius, which is consistent with previous studies~\citep[e.g.,][]{1996AJ....111.1529G,1999ApJ...513..733K,2001AJ....121..210R}.
The [Fe/H] gradient for the combined sample (gray solid line) is best described by
\begin{equation}
[{\rm Fe/H}]=-0.32 \log_{10} (R/{R}_{\rm eff} )-0.76.
\end{equation}
The mean [Fe/H] of M87 GCs decreases faster along the radial sequence than the least-squares fit to the whole sample. 
The stronger GC metallicity gradient of M87 than the other three galaxies is consistent with the literature~\citep[e.g.,][]{2011ApJ...728..116L}.

\vspace{1mm}
\section{Simulations of Globular Clusters}
\label{sec:simulations}

\subsection{Models for Simple Stellar Populations}
\label{sec:SSP}

The simple stellar population simulations in this paper are based on the Yonsei Evolutionary Population Synthesis (YEPS) model~\citep[Paper I;][]{2013ApJS..204....3C,2017ApJ...842...91C}. 
The model used the Yonsei-Yale (Y\textsuperscript{2}) stellar library~\citep{2001ApJS..136..417Y,2002ApJS..143..499K,2003ApJS..144..259Y} to produce all stages of stars from the main sequence to the red giant branch (RGB), as well as HB and post-HB stars. 
Flux libraries are required to obtain the integrated spectral energy distribution of each GC, and our model employed the library of \citet{2002A&A...381..524W}. 
For the initial mass function, our fiducial model used the Salpeter function~\citep{1955ApJ...121..161S}. 
After significant mass loss, RGB stars have a thin hydrogen envelope and further evolve into HB stars that are prominent in blue light. 
Thus, the mass loss is an essential factor in HB modeling. 
We adopted \citeauthor{1977A&A....57..395R}'s \citeyearpar{1977A&A....57..395R} formula for the mass loss during the RGB stage and calibrated the mass-loss rate parameter by comparing the HB morphologies of simple stellar population models to those of observed GCs in the Milky Way.
Readers are referred to~\citet{2013ApJS..204....3C,2017ApJ...842...91C} for a detailed description of the YEPS model including its ingredients and input parameters.

\begin{figure*}[htbp]
\centering
\includegraphics[keepaspectratio,height=0.40\textheight]{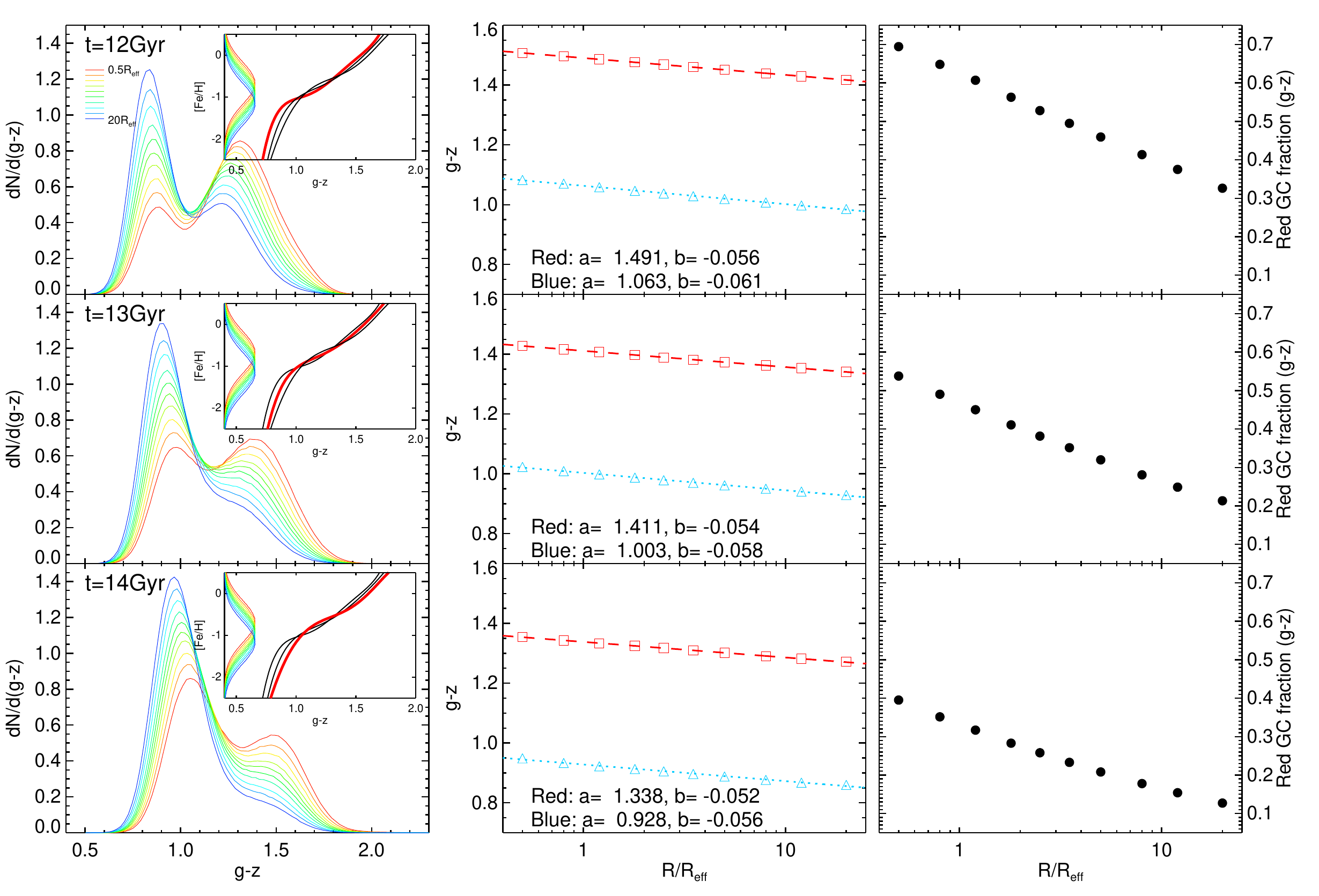}
\caption{\label{fig:Figure5}
(Left column) Radial variations of our model color distributions in $g-z$ color. 
Top, middle, and bottom panels represent the GC color distribution models of 12, 13, and 14 Gyr, respectively. 
The color code shows the radial variation of the mean [Fe/H] of GCs, which is defined in Equation (1) and the radial variation of color distribution functions. 
The insets show the theoretical CMRs, and the red lines represent the pertinent age. 
(Middle column) The radial color gradient of blue and red GCs. 
The blue dotted and red dashed lines are the best linear fit to the mean colors of blue (blue open triangles) and red (red open squares) GCs, respectively. 
The mean colors are calculated by the KMM code and $a$, $b$ are the coefficients in ${g-z} = a+b \log_{10}(R/{\rm R}_{\rm eff})$. 
(Right column) The radial red GC fraction gradient. 
The black circles present the red GC fractions determined by the KMM code.}
\end{figure*}

\subsection{Models for GC Color Distributions}
\label{modelsforgccolordistributions}

Figure~\ref{fig:Figure5} shows how we make the models for GC color distribution morphologies based on the simple stellar population models (Section~\ref{sec:SSP}).
The left column presents the models of GC $g-z$ distributions for different ages (12, 13, and 14 Gyr), where color bimodality is evident. 
As we showed in Figure 1 of Paper VIII, our model suggests that bimodality in color diminishes for stellar populations younger than 12 Gyr.
We use the mean metallicity at a given galactocentric radius (from 0.5\,${R}_{\rm eff}$ to 20\,${R}_{\rm eff}$) based on the empirical metallicity gradient of Equation (1).
At each radial position, we assume a simple Gaussian MDF centered at the mean [Fe/H] value, as presented by the distribution along the y-axes of the insets of the left column.
We convert the MDFs to the model GC color distribution histograms via our theoretical CMRs.
The properties of the color distributions, such as the mean colors of blue and red GCs and the red GC fraction, are acquired by the KMM code assuming the homoscedastic case.
In the middle column, the mean colors of blue and red GCs change along the radii depending on the radial metallicity variation.
As the radial distance increases, the mean colors of blue and red GCs shift toward blue, which is common in observations (see Figure~\ref{fig:Figure2}). 
We will compare in Section~\ref{sec:4.1} our model to the observations in terms of the radial variation in the mean colors of blue and red GCs.
In the right column, the model shows the systematic radial variation in the red GC fraction in the sense that the red fraction increases as the mean metallicity of the MDF increases.

It is noteworthy that for different ages, even with the same MDF, the model color distributions show differences in (a) the mean colors of blue and red GCs and (b) the red GC fraction.
This is because the detailed shape (i.e., the color of the inflected position and the degree of inflection) of the CMRs varies systematically with their ages (see Figure 1 of Paper VIII for more details).
The inflection point of a CMR moves toward a redder color when a model age gets older. 
In addition, the inferred age of a GC system gives the age of the oldest stellar populations in its host galaxy.
Particularly for early-type galaxies, the spectroscopic ages of the central field stellar population of galaxies~\citep[e.g.,][]{2010MNRAS.408...97K} concur with those of their GC systems~\citep[e.g.,][]{1998ApJ...496..808C}.
This notion opens up a new possibility of galaxy age dating by exploiting the observed color distribution morphologies of extragalactic GC systems. 
As pointed out in Paper VIII, this age determination technique is based on photometry rather than spectroscopy, and it has precision as good as $\lesssim$$\pm$1 Gyr for massive galaxies hosting abundant GCs. 
An additional advantage is that the technique is less affected by the notorious age$-$metallicity degeneracy of the stellar population compared to other age-dating methods using galactic integrated colors and spectra (see Paper VIII for details).
To determine the age of GC systems, we construct color distribution models of various ages from 8 to 15 Gyr by 0.1 Gyr intervals.
We will apply this methodology to our sample GC systems and infer their radial age gradient in Sections~\ref{sec:4.1} and~\ref{sec:4.2}.

\section{Results and Analysis}
\label{sec:results}

\subsection{Radial Variation in the GC Color Distribution Morphology}
\label{sec:4.1}

\begin{figure*}[htbp]
\centering
\includegraphics[keepaspectratio,height=0.67\textheight]{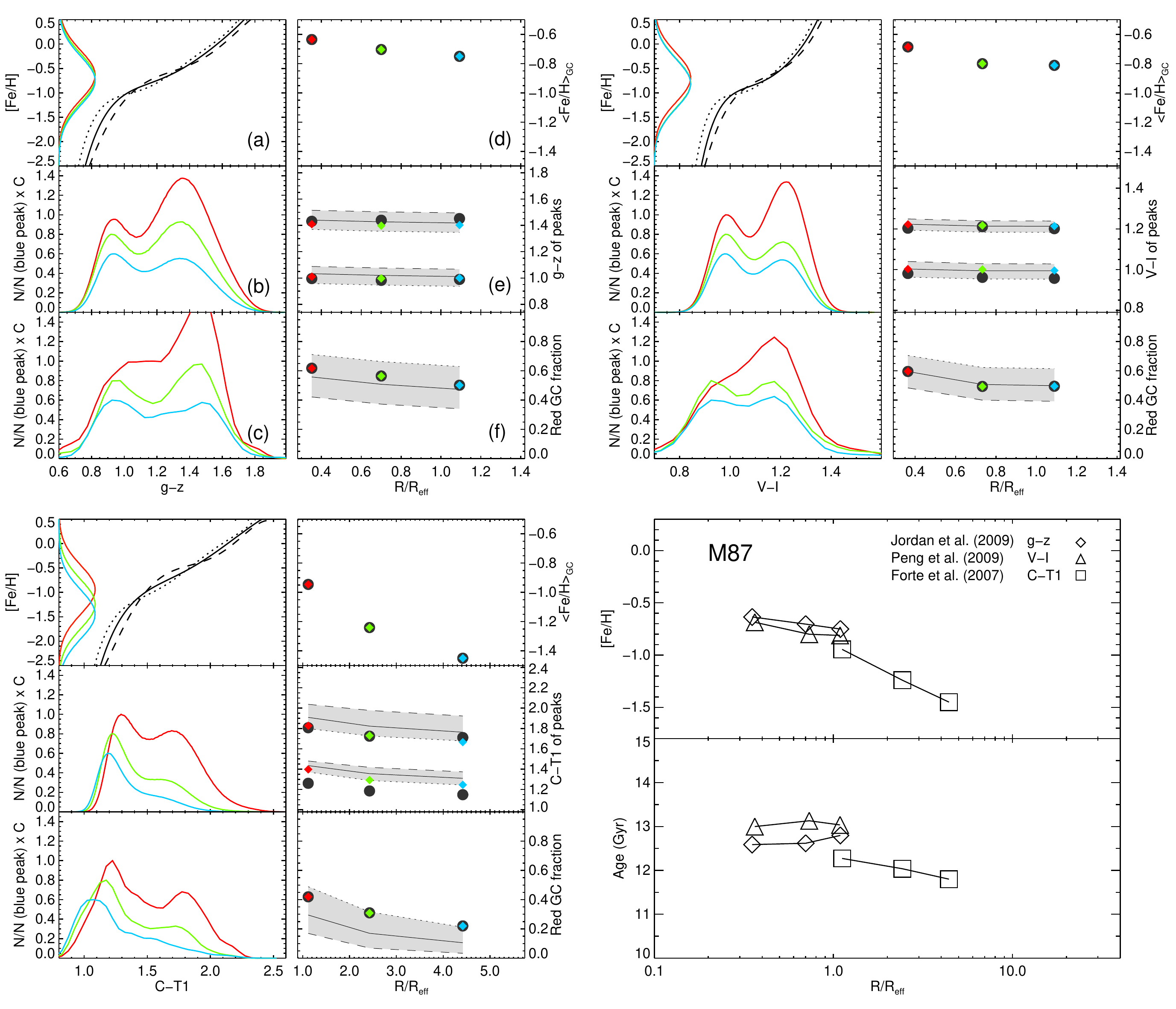}
\caption{\label{fig:Figure6} 
Upper left set of panels: the $g-z$ color distribution of M87 with respect to the radius. The observational data are taken from \citet{2009ApJS..180...54J}.
(a) The black lines present the YEPS color$-$[Fe/H] relations for 12 Gyr (dotted line), 13 Gyr (solid line), and 14 Gyr (dashed line).
Gaussian distributions on the y-axis with red, green, and blue colors show model metallicity distributions.
The values of the mean [Fe/H] of GCs are adopted from the observations shown in (d).
(b) The best-matched color distribution models from comparing with observed red GC fractions shown in (f). 
(c) The observed color distributions with Gaussian kernels with $\sigma$(color) = 0.05.
(d)$-$(f) The black circles are the observed mean [Fe/H] of GCs, the mean colors of blue and red GCs, and the fraction of red GCs as a function of radius. 
The model results are marked by diamonds with the same colors as (a)$-$(f).
The dotted, solid, and dashed lines in (e) and (f) are the isoage models corresponding to the model color$-$[Fe/H] relation in (a).
Upper right set of panels: M87 in $V-I$. 
The observational data are taken from \citet{2009ApJ...703...42P}. 
Lower left set of panels: M87 in $C-T_1$. 
The observational data are taken from \citet{2007MNRAS.382.1947F}. 
Lower right set of panels: radial variation of ${\langle{\rm[Fe/H]}\rangle_{\rm GC}}$ and ${\langle{\rm Age}\rangle_{\rm GC}}$ of M87.}
\end{figure*}

\begin{figure*}[htbp]
\centering
\includegraphics[keepaspectratio,height=0.67\textheight]{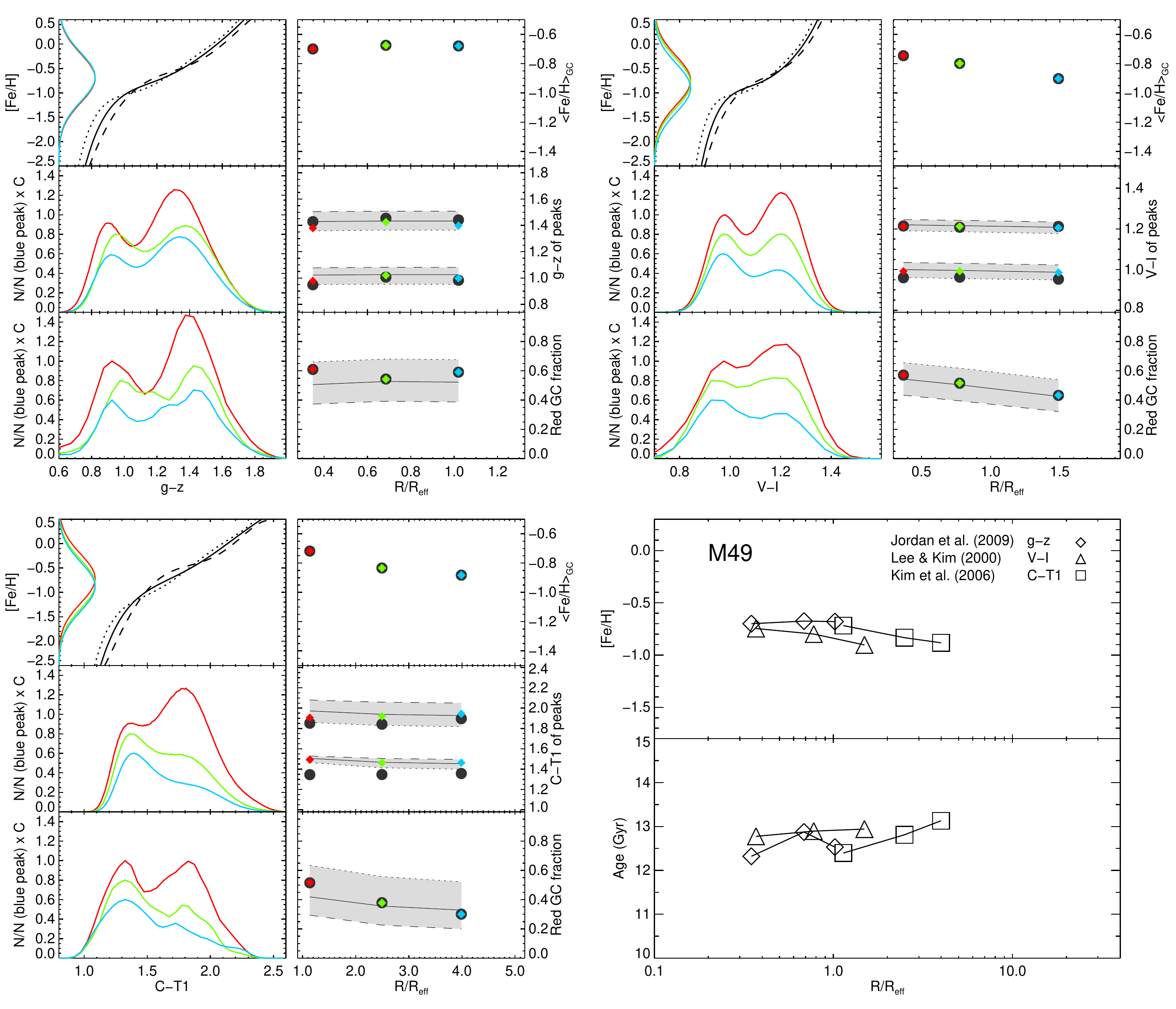}
\caption{\label{fig:Figure7}
Same as Figure~\ref{fig:Figure6}, but for M49. Observational data are taken from~\citealt{2009ApJS..180...54J} (upper left set of panels), \citealt{2000AJ....120..260L} (upper right set of panels), and~\citealt{2006ApJ...647..276K} (lower left set of panels).}
\end{figure*}

\begin{figure*}[htbp]
\centering
\includegraphics[keepaspectratio,height=0.67\textheight]{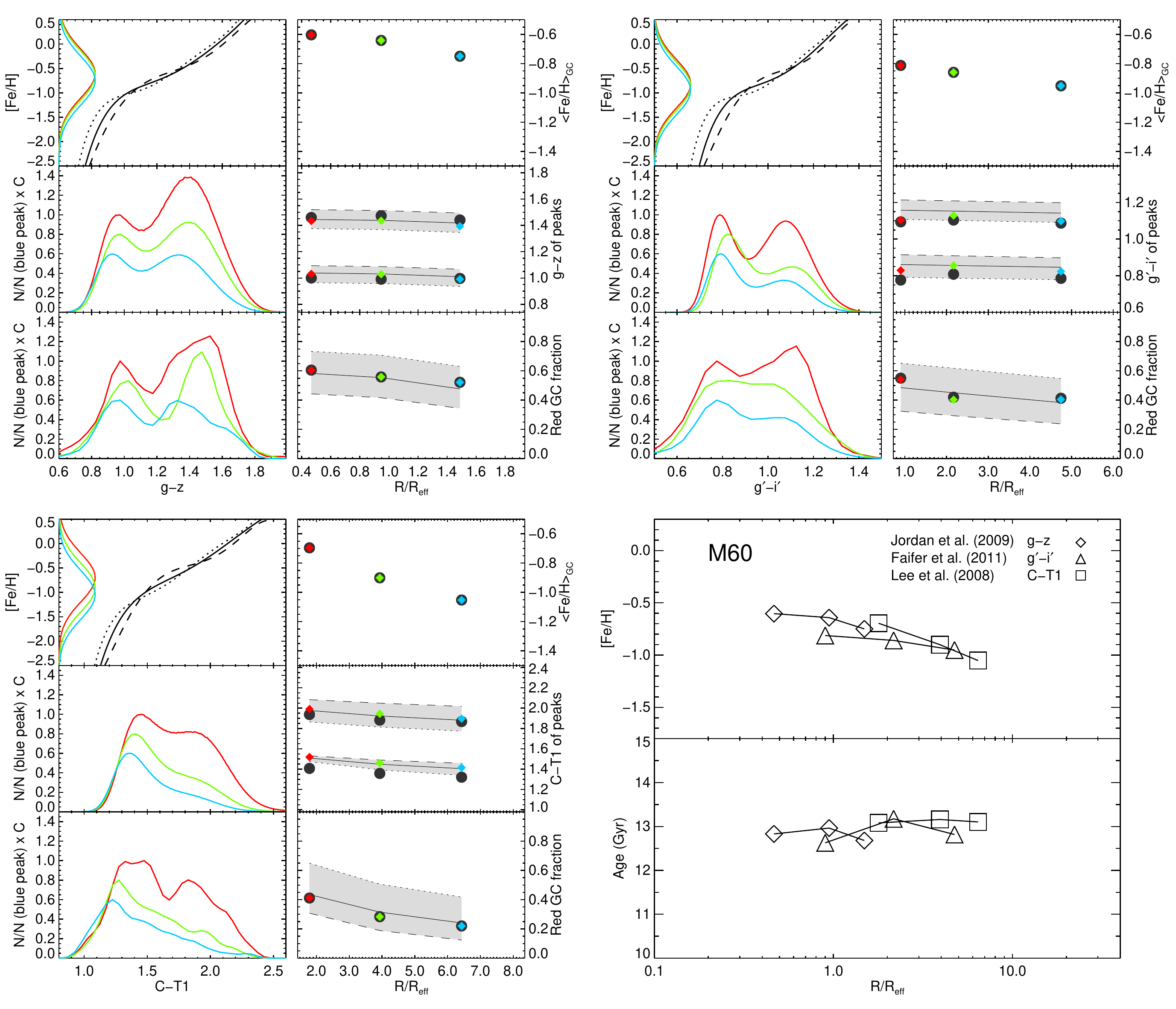}
\caption{\label{fig:Figure8}
Same as Figure~\ref{fig:Figure6}, but for M60. Observational data are taken from~\citealt{2009ApJS..180...54J} (upper left set of panels), \citealt{2011MNRAS.416..155F} (upper right set of panels), and \citealt{2008ApJ...682..135L} (lower left set of panels).}
\end{figure*}

\begin{figure*}[htbp]
\centering
\includegraphics[keepaspectratio,height=0.67\textheight]{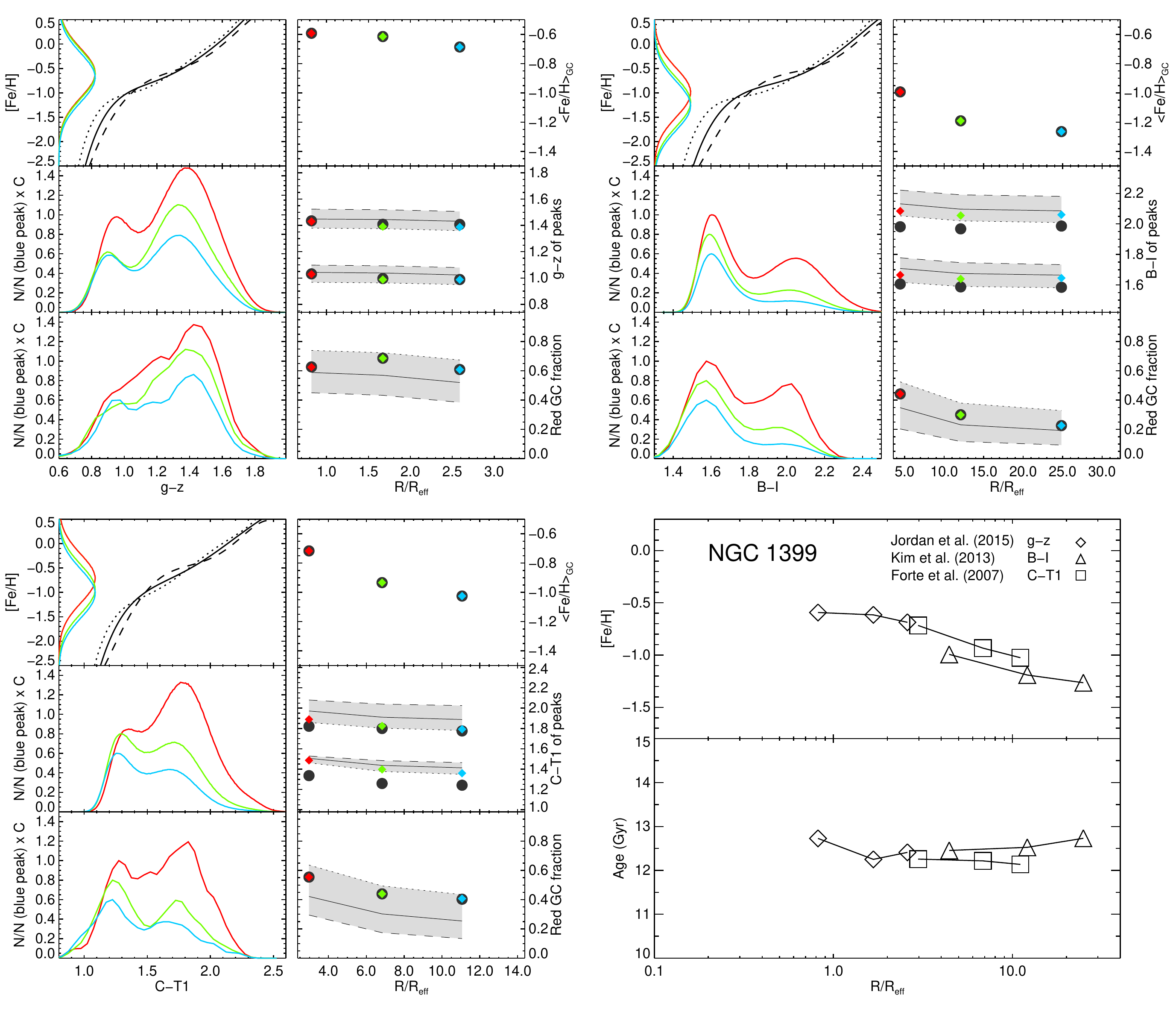}
\caption{\label{fig:Figure9}
Same as Figure~\ref{fig:Figure6}, bur for NGC 1399. Observational data are taken from \citealt{2015ApJS..221...13J} (upper left set of panels), \citealt{2013ApJ...763...40K} (upper right set of panels), and \citealt{2007MNRAS.382.1947F} (lower left set of panels).}
\end{figure*}

\begin{figure*}[htbp]
\centering
\includegraphics[keepaspectratio,height=0.38\textheight]{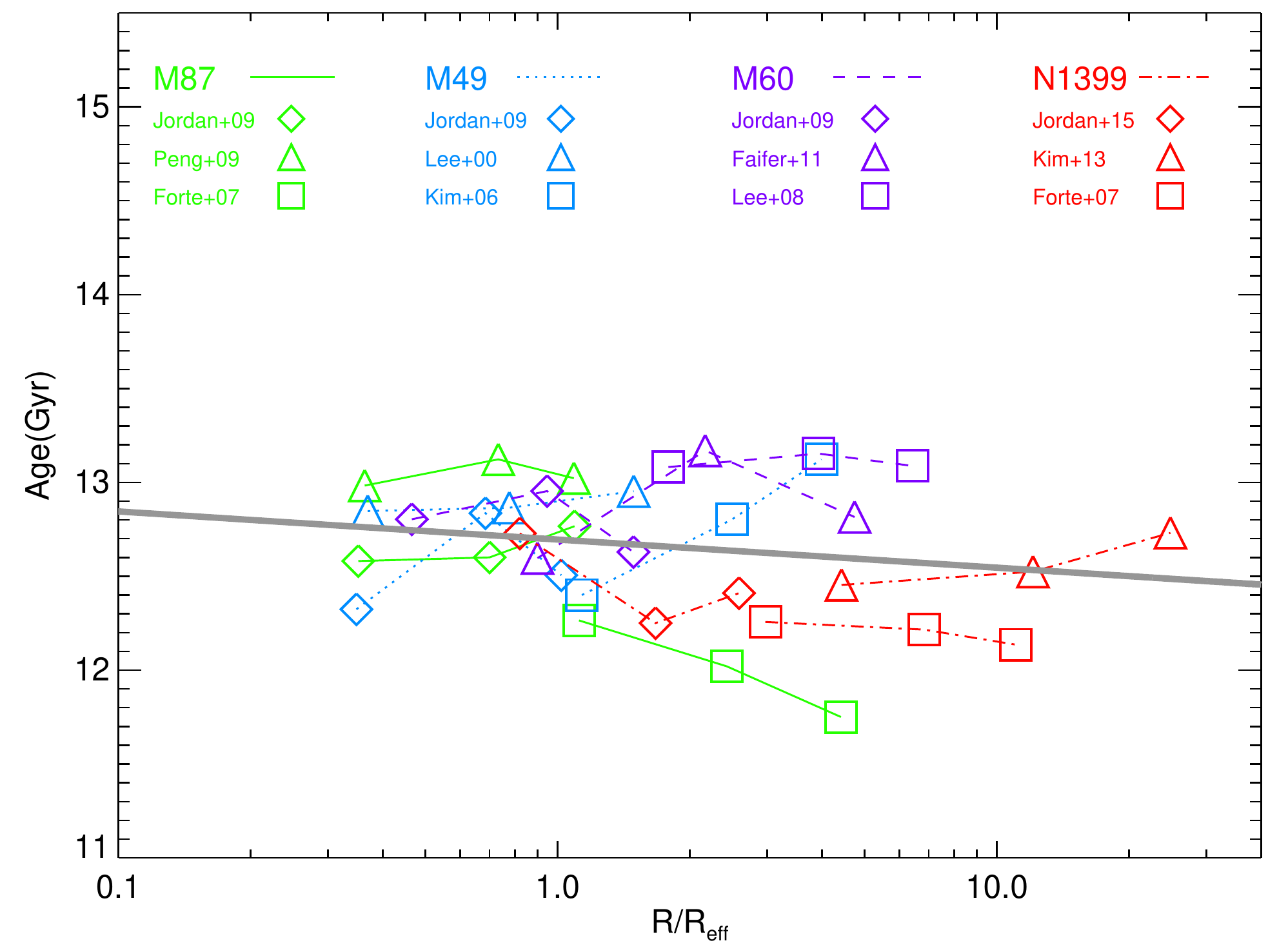}
\caption{\label{fig:Figure10}
Radial variation of the inferred GC ages of our sample galaxies. The gray solid line shows the least-squares fit to the whole data.}
\end{figure*}

In Figures~\ref{fig:Figure6}$-$\ref{fig:Figure9}, for the four sample galaxies, our models are compared with the radial variation of the observational color distribution morphologies.
In Figure~\ref{fig:Figure6}(a), we present our model prediction of CMRs for 12, 13, and 14 Gyr. 
The simple Gaussian MDFs with ${\sigma\rm[Fe/H]_{GC}}=0.5$ are shown along the y-axis.
The mean [Fe/H] of the GC MDFs are derived from colors via the equations in Table~\ref{tab:Table2} and shown in Figure~\ref{fig:Figure6}(d).
Figures~\ref{fig:Figure6}(b) and (c) present the modeled and observed color distributions, respectively.
The model color distributions in Figure~\ref{fig:Figure6}(b) are obtained from the corresponding MDFs (in Figure~\ref{fig:Figure6}(a)) via the nonlinear CMR of the best-matching age. 
The best-matching ages are derived from Figure~\ref{fig:Figure6}(f) by comparing the observed red GC fractions with the model grids.
Remarkably, the inflected CMRs plus the radial GC metallicity gradient reproduce well the systematic radial trend of the GC color distribution morphologies. 
Figures~\ref{fig:Figure6}(e) and (f) show the radial variations in the mean colors of blue and red GCs and in the red GC fraction, respectively.
The observed mean colors of blue and red GCs and the observed red GC fraction are placed well within our model grids of ${\langle\rm Age \rangle_{GC}}=$ 12$-$14 Gyr.
The top-right and bottom-left sets of panels are for the other observational datasets using different colors and covering different radial zones. 
The bottom-right set of panels shows the mean [Fe/H] (upper) and the inferred age (lower) as functions of the radius of the host galaxy.

The format of Figure~\ref{fig:Figure6} is repeated for Figures~\ref{fig:Figure7}$-$\ref{fig:Figure9}.
In what follows, we make comments on the result for each galaxy (Figures~\ref{fig:Figure6}$-$\ref{fig:Figure9} each). 
In Figure~\ref{fig:Figure6}, for M87, the red GC fractions of the three observational catalogs suggest the age of the GC system in all radial bins to be between 12 and 13 Gyr.
The first two data sets with a similar radial coverage use different filter systems ($g-z$ and $V-I$); however, the mean age estimations are entirely consistent with each other.
The observed field in $C-T_1$ is a more outer region than those of the two other observations, yet the inferred age is similar to the ages from the other two. 
The mean colors of the blue GCs show a reasonable agreement between the observations and models.
For $C-T_1$, however, the mean colors of blue GCs of the models are redder by 0.1$-$0.2 mag than those of the observations.

In Figure~\ref{fig:Figure7}, for M49, the observed red GC fractions and mean color of blue and red GCs in $g-z$ and $V-I$ agree well with our model grids of ${\langle\rm Age \rangle_{GC}}$ = 12$-$13 Gyr.
The observational color histograms in $g-z$ show more irregular shapes than other observations due to their small numbers.
The mean colors of blue GCs of the models are redder than the observation by $\sim$\,0.1 mag in $C-T_1$. 
In Figure~\ref{fig:Figure8}, for M60, the observed $g-z$ distributions show the feature of small numbers.
The ages from $g-z$ and $g^{\prime}-i^{\prime}$ colors are well within our 12$-$13 Gyr model grids.
For the $C-T_1$ color, the absolute values of ages are slightly older than the two other cases.
In Figure~\ref{fig:Figure9}, for NGC 1399, the comparison of the red GC fractions between the observations and models indicates that the age variation along the radius is negligible.
The mean colors of blue and red GCs in $C-T_1$ do not match well with our model grids (12$-$14 Gyr).
Generally, our models related to the short-wave bands ($C$ and $B$ bands) predict slightly redder peaks than the observations. 
This may be due to the still incomplete modeling for hot HB and post-HB stars.

The radial gradient of the mean colors of blue and red GCs is a controversial issue (see Figure~\ref{fig:Figure2}).
A number of observations~\citep{1996AJ....111.1529G,2009ApJ...699..254H,2009ApJ...703..939H,2011MNRAS.413.2943F,2011ApJ...728..116L,2012MNRAS.421..635F,2018MNRAS.479.4760F} reported that the mean colors of both blue and red GCs get bluer with galactocentric radius. 
Some found the radial gradient for either red or blue group~\citep{2001AJ....121.1992F,2011ApJS..197...33S,2016ApJ...817...58H,2017MNRAS.470.3227C} or no gradient for both groups~\citep{2009AJ....137.3314H,2017ApJ...835..101H}.
Back in Figure~\ref{fig:Figure5}, our model shows shallow radial color gradients for both blue and red GCs in optical $g-z$ colors.
In the case of the 13 Gyr model, when we adopt the [Fe/H] gradient of our sample galaxies (Equation (1)), the $g-z$ color gradients (${\Delta (g-z)}/{\Delta \log R}$) of the blue and red GCs are $-0.06$ and $-0.05$, respectively, for 0.5$-$20\,${R}_{\rm eff}$.
\citet{2011ApJ...728..116L} analyzed GC systems in 76 early-type galaxies in the ACS Virgo Cluster Survey and Fornax Cluster Survey data.
Excluding those classified as unimodal distributions, they found the mean $g-z$ color gradient of blue and red GCs in their 39 galaxies to be $-0.04 \pm 0.01$ and $-0.05 \pm 0.01$, respectively.   
For the four giant ellipticals (M87, M49, M60, and NGC 1399), the mean values of the blue and red GCs gradients are $-0.04 \pm 0.01$ and $-0.04 \pm 0.01$, respectively. 
Considering the difference\footnote{\citet{2011ApJ...728..116L} divided GCs in the entire radial range into the blue and red GCs using a simple color cut determined by the KMM test, while we perform the KMM test for every radial bin. See~\citet{2011ApJ...728..116L} for more details. See also~\citet{2020ApJ...900...95V} for an explanation that the simple color cut can bias the gradient measurement.} in measuring the gradient, our model prediction agrees well with the result of~\citet{2011ApJ...728..116L}.

\subsection{Radial Gradient of the GC Ages}
\label{sec:4.2}

The inflection point of a CMR is located at a redder color for an older age model. 
As a consequence, for a given MDF, an older CMR produces a lower red GC fraction (see Figure~\ref{fig:Figure5}).
This is the principle behind the age dating based on the GC color distribution morphology.
In this regard, we find that the red GC fraction is a more robust age indicator than the mean colors of blue and red GCs, especially for the short-wave band colors (e.g., $C-T_1$) that are more vulnerable to the incompleteness of the model ingredients (Section 4.1).
We thus prefer using the former over the latter.
It is fair to note that the mean metallicities of GC systems may be affected by the detailed shape of CMRs that is still somewhat uncertain.
With overestimated (underestimated) mean [Fe/H], the age inferred from the red GC fraction would be higher (lower) than the \textit{true} age.

In Figure~\ref{fig:Figure10}, we show the inferred ages of the GC systems as a function of the radius normalized by ${R}_{\rm eff}$.
This is a combined figure of the lower panels of the bottom-right sets of panels in Figures~\ref{fig:Figure6}$-$\ref{fig:Figure9}.
The inferred ages of the GC systems are distributed around 13 Gyr and do not show a significant radial gradient.
The age variation within each galaxy is as low as 1 Gyr. 
No evidence of the radial age gradient implies that the radial variation in the color distribution morphology arises predominantly due to the mean [Fe/H] gradient rather than due to the age gradient.

\begin{table*}
\footnotesize
\begin{center}
\caption{The Number Density Profiles of Total GCs and the Empirical [Fe/H]$-$Radius Relations\label{tab:Table3}}
\begin{tabular}{lccccccccccccccclcccccccccccc}
\tableline
\tableline
 \  Galaxy & & & & & & \multicolumn{11}{c}{$\Sigma=\alpha+\beta(R/{R}_{\rm eff})^{1/4} $}& & & & & & &\multicolumn{6}{c}{$[{\rm Fe/H}] =\delta+\gamma(R/{R}_{\rm eff})^{1/4}$}\\
\cline{7-17}
\cline{24-29}
           & & & & & & $\alpha$ & & & & & $\beta$  & & & & & Data Source                     & & & & & & &$\delta$   & & & & & $\gamma$\\ 
\tableline
       M87 & & & & & & 3.821    & & & & & $-$1.868 & & & & & \citet{2006MNRAS.373..601T}     & & & & & & & 0.350     & & & & & $-$1.220\\
       M49 & & & & & & 2.773    & & & & & $-$1.415 & & & & & \citet{1998AJ....115..947L}     & & & & & & & $-$0.477  & & & & & $-$0.287\\
       M60 & & & & & & 3.350    & & & & & $-$1.917 & & & & & \citet{2008ApJ...682..135L}     & & & & & & & $-$0.173  & & & & & $-$0.532\\
  NGC 1399 & & & & & & 2.401    & & & & & $-$1.358 & & & & & \citet{cantiello18}             & & & & & & & 0.002     & & & & & $-$0.588\\
\tableline
\end{tabular}
\end{center}
\tablecomments{The empirical [Fe/H]$-$radius relations are fitted to the mean [Fe/H] of our sample datasets binned into three radial regions for each sample.}
\end{table*}

\subsection{Radial Number Density Profiles of Blue and Red GCs}
\label{sec:4.3}

\begin{figure*}[htbp]
\centering
\includegraphics[keepaspectratio,width=\textwidth,height=0.87\textheight]{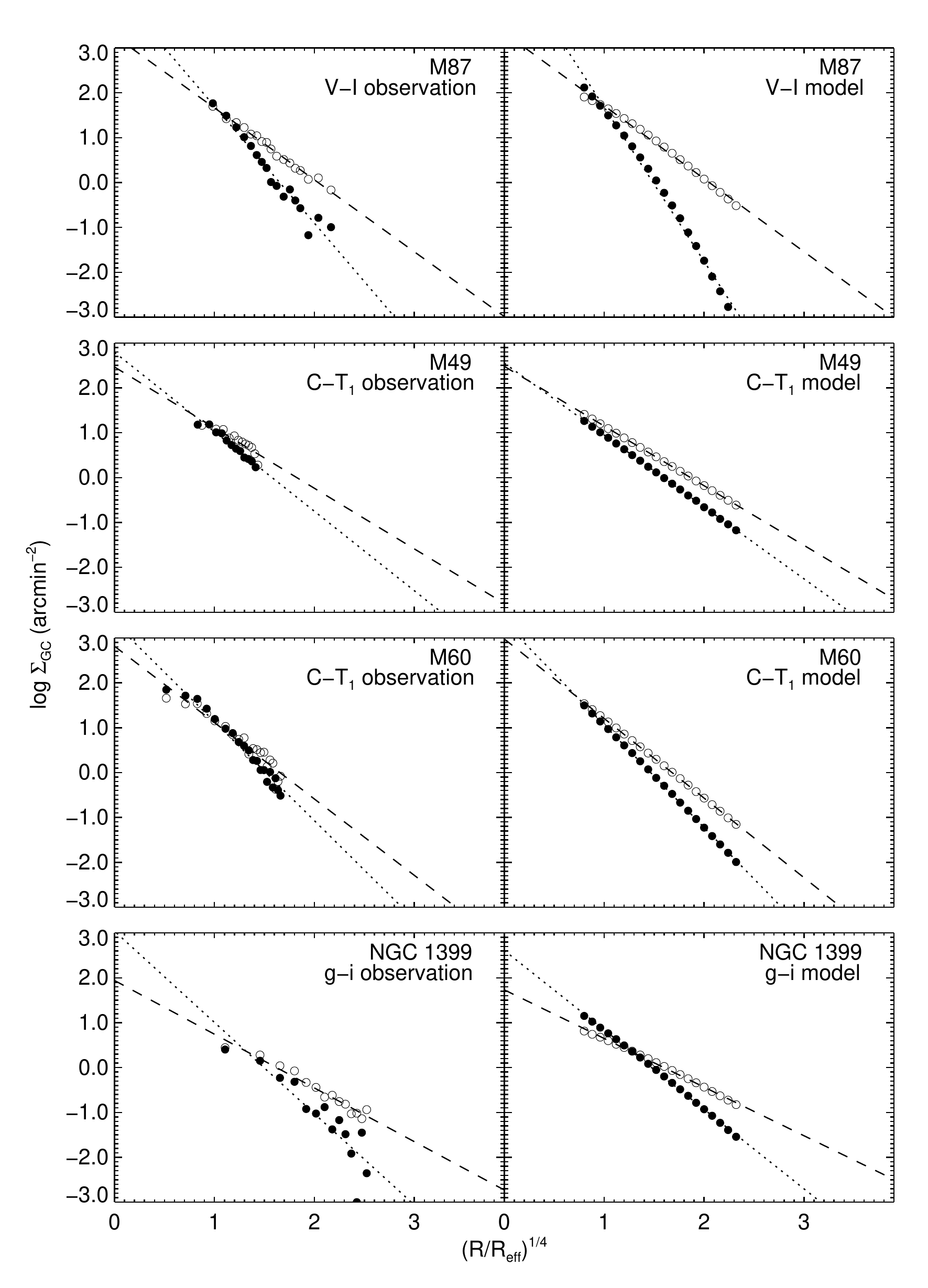}
\caption{\label{fig:Figure11}
Comparison between observations (left column) and models (right) of the surface number density profiles.
The open and filled circles are the number density of blue and red GCs, respectively. 
The dashed and dotted lines are the least-squares fit to the blue and red GCs, respectively.
The observational number density profiles (left column) are the same as Figure~\ref{fig:Figure1}.
The model number density profiles (right column) are calculated by combining (a) the modeled radial profile of the number ratio of blue and red GCs and (b) the observed radial profile of the total (blue + red) GC number density (M87:~\citealt{2006MNRAS.373..601T}; M49:~\citealt{1998AJ....115..947L}; M60:~\citealt{2008ApJ...682..135L}; NGC 1399:~\citealt{cantiello18}).
} 
\end{figure*}

Figure~\ref{fig:Figure11} compares our models for the radial surface number density profiles of blue and red GCs with the observations.
The radial profile models are generated by combining (a) the radial variation in the GC color distribution morphologies depending on the observed [Fe/H] gradient of GCs (as shown in Section~\ref{sec:4.1}) and (b) the observed radial number density gradient of the entire (blue + red) GCs.
Table~\ref{tab:Table3} gives the observed [Fe/H] gradient and the observed surface number density gradients. 
The age of the GC systems is assumed to be 13 Gyr.
In the left column, the observed number density profiles show that the blue and red GCs have different slopes.
Conventionally, the difference in the radial density profile between blue and red GCs has been ascribed to their distinct origins and spatial distributions~\citep[e.g.,][]{2004MNRAS.355..608F,2011ApJS..197...33S}.

In the right column, however, our model reproduces the different slope of the number density profiles of the blue and red GCs naturally.
Our model predictions show that the intersecting points of the two profiles are located at the slightly more inner region of galaxies than those of observations.
The difference seems due to the uncertainty of the empirical [Fe/H]$-$color relations.
We also note that the simulated difference in the slope between blue and red GCs for each galaxy slightly differs from the observation.
According to our simulation, even if the total (blue + red) number density profile of GCs is the same, the steeper the metallicity gradient is, the more significant the difference in the slope between the blue and red GCs is.
The remarkable agreement between our models and observations leads us to conclude that the difference in the radial profile between blue and red GCs stems naturally from the combined effect of the radial metallicity gradient of GC systems plus the metallicity-to-color nonlinearity.
We, therefore, propose that there is no need for assuming the distinct origins of the blue and red GCs to explain their observed difference in the radial surface number density profile.

\vspace{1mm}

\section{DISCUSSION}
\label{sec:discussion}

We have demonstrated that our theoretical model reproduces the observed radial variations in terms of (a) the GC color distribution morphologies such as relative portions of blue and red GCs and (b) the surface number density of blue and red GCs. Our results provide an alternative, more cohesive solution to the distinct radial density profiles of blue and red GCs that does not necessarily invoke two GC subsystems and thus reinforces the nonlinear-CMR scenario for the GC color bimodality.

Our simulation shows that the radial variation in the ensemble average of GC ages is within 1 Gyr out to the radial range of $\sim$20\,${R}_{\rm eff}$.
This implies that GCs throughout the wide radial extent were created in a coeval manner for our giant elliptical galaxies.
A number of studies reported that the radial age gradient of field stars in early-type galaxies is almost flat~\citep{2003A&A...407..423M,2006MNRAS.369..497K,2006A&A...457..823S,2007MNRAS.377..759S,2008MNRAS.385..675S,2014MNRAS.439..990M} with a negative metallicity gradient~\citep[e.g.,][]{2020ApJ...896...75S}.
The inferred age distributions of field stars are based on long-slit and integral field unit spectroscopy and mostly confined to the central region of galaxies ($\lesssim$2\,${R}_{\rm eff}$).
By contrast, our methodology using the GC color distribution allows us to investigate the age variation of the wide range ($\sim$20\,${R}_{\rm eff}$).
No or little radial gradient in ages of both GCs and field halo stars suggests that GC systems and their parent galaxies have shared a more common history than previously thought (see Paper III).

Several studies have reported that observational properties of unresolved halo field stars, such as the surface light profile~\citep{2008MNRAS.386.1145B,2012MNRAS.425...66F,2014ApJ...794..103D,2018MNRAS.474.4302E}, ellipticity~\citep{2013ApJ...773L..27P,2014MNRAS.437..273K}, and kinematics~\citep{2010A&A...513A..52S,2011ApJS..197...33S, 2013MNRAS.428..389P,2020A&A...637A..26F}, are often more similar to those of red GCs than blue GCs. 
These findings have been regarded as the clues that there are two distinct GC subpopulations and the red subpopulation shares a more intimate history with field stars of a host galaxy. 
On the contrary, our nonlinearity scenario suggests that many thousands of building blocks were involved in making a single massive galaxy. 
Such a notion leaves little room for the existence of just two GC groups in individual galaxies. 
In this regard, Paper III showed that the true MDF of GCs is of a unimodal, skewed Gaussian shape with a metal-poor tail, similarly to that of field stars, suggesting that both GCs and field stars underwent a continuous chemical enrichment with a short timescale (2\,$\sim$\,3 Gyr).
The properties of field stars and GCs were developed simultaneously over such a timescale. 
Recently, \citet{2020ApJ...900...95V} showed that blue and red GCs as well as halo field stars in M87 have a similar metallicity gradient in the inner halo ($<$40 kpc), further supporting a common origin of GCs and halo field stars of galaxies.
From our point of view, the observed mean properties of field stars are better represented by metal-rich stars that occupy the majority of field stars at and around the peak of their MDFs. 
It is the red GC group that has a metallicity value similar to metal-rich field stars. 
Thus, naturally, the red GCs better follow the properties of unresolved field stars than the blue GCs.

As described in our previous papers (Papers II, III, and IV), even with identical MDFs, the color distribution morphology would vary depending on colors used in observations.
This is because the exact CMR shape depends on the color.
Given such color-dependent variation of GC color distributions, the radial density profiles of blue and red GCs should vary when different colors are used, even for the same galaxy. 
\citet{2014ApJ...794..103D} obtained the number density profiles of blue and red GCs in M87 and compared their result (from $g\arcmin-i\arcmin$) with the result (from $V-I$) of \citet{2006MNRAS.373..601T} for the same galaxy.
The blue GC profiles of the two studies showed a similar slope, but the red GC profiles were different. 
Durrell et al. ascribed the discrepancy to the different criteria for dividing blue and red GCs. 
However, we suspect that the used colors also affect the difference between the two results.
In this vein, the variation of the number density profiles due to the change of observed colors can be estimated based on multiband photometry of GCs. 
For instance, \citet{2013ApJ...763...40K} provided $UBVI$ colors of the GC systems in the Fornax galaxy cluster.
In their Figure 15, the red GC fraction in $U-B$ shows a different gradient from that in other colors for NGC 1399. 
A clearer difference in the density profile according to the different choice of colors is expected for the optical/NIR color combination.
Several optical/NIR studies reported that the number fractions of blue and red GCs vary with colors~\citep{2012ApJ...746...88B,2012A&A...539A..54C,2016ApJ...822...95C}.
Thus, with the upcoming James Webb Space Telescope, the high-S/N NIR data on a large number of GC systems will be crucial to test our explanation for the different radial density profiles of blue and red GCs as well as the color bimodality itself.

\vspace{1mm}
S.-J.Y. acknowledges support by the Mid-career Researcher Program (No. 2019R1A2C3006242) and the SRC Program (the Center for Galaxy Evolution Research; No. 2017R1A5A1070354) through the National Research Foundation of Korea. The updated data ($CT_1$) for NGC 4472 are kindly provided by Eunhyeuk Kim.

\bibliographystyle{aasjournal}

\end{document}